\documentclass[fleqn,twoside]{article} \usepackage{espcrc2}

\usepackage{graphicx}
\usepackage[figuresright]{rotating}

\newcommand{\AmS}{{\protect\the\textfont2
A\kern-.1667em\lower.5ex\hbox{M}\kern-.125emS}}

\title{The mass of the charm quark from unquenched lattice QCD at
fixed lattice spacing.}

\author{UKQCD Collaboration\\
	 A. Dougall\address{School of Mathematics,
        Trinity College, Dublin 2, Ireland},
        C. M. Maynard\address{School of Physics, University of
        Edinburgh, Edinburgh EH9 3JZ, UK},
        C. McNeile\address{Theoretical Physics, Dept. of Mathematical
        Sciences, University of Liverpool, Liverpool L69 3BX, UK}}
       
\begin{document}

\begin{abstract}
We determine the mass of the charm quark ($m_c$) from lattice QCD with two
flavors of dynamical quarks with a mass around the strange quark. We 
compare this to a determination in quenched QCD which has the same lattice 
spacing (0.1 fm). We investigate different formulations of the quark mass,
based on the Vector Ward Identity, PCAC relation and the FNAL heavy quark 
formalism. Based on these preliminary results we find no effects due to
sea quarks with a mass around strange.

\vspace{1pc}
\end{abstract}

\maketitle

\section{INTRODUCTION}
Quark masses are fundamental parameters of the Standard Model but due
to confinement they cannot be measured directly by experiment and
therefore any mass quoted always depends on the scheme (and scale) in
which it was computed. The present world average value for
$m_c^{\overline{\mathrm{MS}}}(m_c)$ has an error of approximately
17\%~\cite{PDG2002}.  This should be compared to the value quoted for
the experimental mass of the $D$ meson which is accurate to within
less than 0.1\%.

The mass of the charm quark is a difficult quantity to compute since it 
is light enough to be challenging for heavy quark methods
and yet heavy enough such that the lattice spacing
may be too coarse for the Compton wavelength. There has recently been
several calculations of this quantity from the lattice
\cite{deD,Rolf,Bec,Juge}. In particular, a detailed study by Rolf and
Sint \cite{Rolf} has been presented in the continuum limit, thus
eliminating the error arising from lattice artifacts. However, the
effects of including quark loops within these calculations has so far
been omitted. This work is an attempt to estimate these effects and
obtain a result in full QCD (albeit at fixed lattice spacing). In the
interest of reducing errors resulting from a finite lattice spacing,
we investigate several different formulations of $m_c$ on the lattice.

We have previously published results from this data set on the
hyperfine and P-wave mass splittings~\cite{splitting}. The mass
splittings between the P-wave and S-wave mesons have been compared to
the mass of the $D_{sJ}^*(2317)^+$ meson recently discovered by BaBar.

\section{DEFINING THE QUARK MASS ON THE LATTICE}

In this section we outline the definitions of bare quark mass used in
this study and discuss some aspects of the renormalisation and
improvement procedures. We use three different definitions, the first of 
which is the vector Ward Identity (VWI), given by
\begin{eqnarray}
m_{V}=\frac{1}{2}\left( \frac{1}{\kappa_h}-\frac{1}{\kappa_{\rm crit}} \right).
\end{eqnarray}
where $\kappa$ is the hopping parameter, the subscript $h$ denotes the
heavy quark and the subscript crit denotes the value of $\kappa$ which
corresponds to zero quark mass.

The second definition arises from the PCAC relation (AVWI) and is
given by
\begin{eqnarray}
m_{A}^{Q} + m_{A}^{q} =\frac{\langle \sum_x \partial_4 A^I_4(x)
P^{\dagger} (0)\rangle}{\langle \sum_x P(x)P^{\dagger}(0) \rangle}
\end{eqnarray}
where $Q$ and $q$ label the heavy and light quark respectively, $A$ is
the local axial current and $P$ is the pseudoscalar density.  The
axial current has been ${\mathcal O}(a)$ improved.

The scaling studies of Rolf and Sint~\cite{Rolf} show that  ${\mathcal
O}(a^2)$ lattice artifacts of the non-perturbative renormalisation
scheme can be large. As some of the required coefficients have not
been computed non-perturbatively at the parameter values used in this study,
we use the (boosted) perturbative values, thus the lattice
artifacts are ${\mathcal O}(\alpha_s a)$. As the  lattice spacing is
relatively coarse, we utilise an alternative definition of $m_c$. This
is obtained by using the bare value obtained from the VWI, which is
then identified as the ``rest mass'' in the FNAL
formalism~\cite{kron}. At tree level, this is defined as
\begin{eqnarray}
am_1 = \log(1+am_V).
\end{eqnarray}
We use the one loop expression to connect the vector mass to the
quark mass from Ref.~\cite{selfenergy}.
In this work we use the rest mass of the hadron to compare with
experiment, in the FNAL formalism the kinetic mass of the hadron
is used. This is planned in future work.

Matching onto the $\overline{\mathrm{MS}}$ scheme is performed at one
loop in perturbation theory for all definitions of the quark mass
at the scale $q=1/a$ and the associated systematic uncertainty is
estimated by matching at $q=\pi/a$.

\section{DETAILS OF THE COMPUTATION}

\subsection{Lattice specifications}
There are four ensembles of gauge configurations available for
computing the meson correlation functions but here we present details
and results for the dynamical and coarsest quenched set only. The
dynamical ensemble was generated with two degenerate flavors of sea
quarks with a mass around the strange quark. Furthermore, the lattice
spacing for the dynamical set was matched to that of the coarsest
quenched ensemble (presented here) to facilitate the study of effects due to
sea quarks without ambiguities arising from different lattice spacing
errors. All ensembles were generated using a Wilson gauge action and
the non-perturbatively ${\mathcal O}(a)$ improved fermion action was
used to generate the quark propagators. These details are summarised
in Table~\ref{table:1} and further details of the procedures for
generating the dynamical ensemble and matching to the quenched data
set can be found in Ref.~\cite{alton}.

\begin{table}[h]\caption{Lattice parameters}
\label{table:1}
\begin{tabular}{cccc}
\hline $(\beta,\kappa_{\mathrm{sea}})$ & volume & $a^{-1}$ (GeV)&
$N_{\mathrm{configs}}$ \\  \hline \hline  
(5.93,0)& $16^3\times32$  & $1.664(^{21}_{24})$ & 347 \\ 
(5.2,0.1350)& $16^3\times32$ & $1.716(44)$ &395 \\  \hline \hline
\end{tabular}\\
\end{table}
\vspace{-0.2cm} The scale has been set using the ``Method of Planes''
technique which corresponds to setting $r_0=0.55\,$fm as the central
value. The associated systematic error is then estimated using
$r_0=0.5\,$fm.

\subsection{General approach}

Having computed the correlators, the meson mass is obtained as a
function of quark mass as follows. We fit to a 2 by 2 matrix
of correlators (elements of which are combinations of 
local and fuzzed operators) and the fit region is obtained
by studying the effective mass plots and monitoring $\chi^2/dof$.

By extrapolating (interpolating) $m_{\rm light}$ to $m_u$ ($m_s$), the $D$
and $D_s$ meson masses are obtained as functions of $m_{\rm heavy}$. We
then plot the data using the following interpolation formula
\begin{eqnarray}
aM_H &=& a_0 + a_1 m_{\rm heavy}
\end{eqnarray}
\noindent
The experimental results for $M_D$ and $M_{D_s}$ are then used to read
off the bare value of $m_c$.

\section{PRELIMINARY RESULTS}
\begin{figure}[htb]
\includegraphics[scale=0.3]{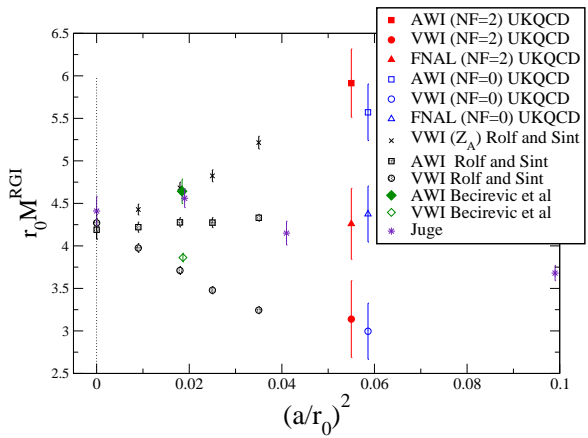}
\caption{Recent data for the charm mass in the RGI scheme as a
function of the lattice spacing.}
\label{worlddata}
\end{figure}

Table~\ref{table:2} contains preliminary results for $m_c(m_c)$ in the
$\overline{\mathrm{MS}}$ scheme for the quenched and partially
quenched sets with the same lattice spacing. The FNAL result presented
here is obtained from the rest mass definition and the errors quoted
are statistical and systematic respectively.  These results have been
converted into Renormalisation Group Invariant (RGI) masses and
plotted against the lattice spacing in Fig.~\ref{worlddata} with some
previous results shown for comparison.
\vspace{-0.3cm}
\begin{table}[htb]
\caption{Preliminary results for $m_c^{\overline{\mathrm{MS}}}(m_c)$.}
\label{table:2}
\begin{tabular}{ccl}
\hline definition & NF & $m_c(m_c)$ (GeV) \\  \hline \hline  AVWI & 2
& 1.705(5)(115) \\  \hline VWI & 2 &  0.905(5)(130) \\  \hline  FNAL &
2 & 1.228(3)(120) \\  \hline AVWI & 0 & 1.606(4)\ (96) \\  \hline  VWI
& 0 & 0.864(5)\ (95) \\  \hline  FNAL & 0 & 1.261(4)\ (95) \\  \hline
\end{tabular}\\
\end{table}

\vspace{-0.6cm}
\section{CONCLUSIONS}
There is no clear evidence of the effects of the sea quarks on the
mass of the charm quark. This is almost certainly due to the
unphysically heavy sea quark masses. However, large systematic
uncertainties make it difficult to compare the quenched and dynamical
results. One major hurdle in reducing the systematic error is the size
of the lattice spacing. The dynamical results have been obtained at a
fixed lattice spacing which is coarse. In comparison with the quenched case,
it is very expensive to reduce the spacing in unquenched simulations
and therefore we are unable to take the continuum limit which would
enable us to control the systematic error. In response to this, we are
investigating which formulation of $m_c$ on the lattice minimizes
discretization errors.

The mass independent renormalisation scheme suffers from large lattice
artefacts of ${\mathcal O}((am)^2)$ for quark masses at coarse lattice
spacings~\cite{Rolf}. Some of the coefficients required for this
scheme have large ${\mathcal O}(a)$ ambiguities themselves. Moreover,
not all of these coefficients are known non-perturbatively for the
dynamical case and so we have resorted to one-loop perturbation theory
in this scheme. The only way to control errors further is to take the
continuum limit which, as discussed, is unfeasible with the current
generation of computers. An alternative approach is that of heavy quark
methods such as the FNAL formalism, which removes mass dependent
lattice artefacts of ${\mathcal O} (\alpha_s (am)^n)$. With careful
application of this formalism, we may be able to make a comparison
between the quenched and dynamical case. This is the subject of future
work, as well as taking the continuum limit in the quenched case to
control remaining lattice artefacts.

\end{document}